\date{March 25, 2024}
\documentclass[letterpaper]{article} % DO NOT CHANGE THIS
\usepackage{arxiv}
\usepackage[utf8]{inputenc} % allow utf-8 input
\usepackage[T1]{fontenc}    % use 8-bit T1 fonts
\usepackage{hyperref}       % hyperlinks
\usepackage{url}            % simple URL typesetting
\usepackage{booktabs}       % professional-quality tables
\usepackage{amsfonts}       % blackboard math symbols
\usepackage{nicefrac}       % compact symbols for 1/2, etc.
\usepackage{microtype}      % microtypography
\usepackage{graphicx}
\usepackage{natbib}
\usepackage{doi}
\usepackage{caption}
\usepackage{color}
\newcommand{\anonymize}[2]{#1}
\usepackage{xcolor}

\usepackage{tabularx}
\usepackage{booktabs}
\begin{document}
\title{Rank, Pack, or Approve: Voting Methods in Participatory Budgeting}

\title{Rank, Pack, or Approve: Voting Methods in Participatory Budgeting}
\author{ \href{https://orcid.org/0000-0003-1936-638X}{\includegraphics[scale=0.06]{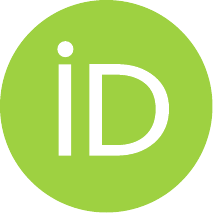}\hspace{1mm}Lodewijk L. Gelauff}\\
    % \thanks{Use footnote for providing further
		% information about author (webpage, alternative
		% address)---\emph{not} for acknowledging funding agencies.} \\
	Center on Democracy, Development and the Rule of Law\\
	Stanford University\\
	\texttt{lodewijk@stanford.edu} \\
	%% examples of more authors
	\And
	% \href{https://orcid.org/0000-0000-0000-0000}{\includegraphics[scale=0.06]{orcid.pdf}
    \hspace{1mm}Ashish Goel\\
	Management Science and Engineering\\
	Stanford University\\
	\texttt{ashishg@stanford.edu} \\
}

\maketitle

\begin{abstract}
Participatory budgeting is a popular method to engage residents in budgeting decisions by local governments. The \anonymize{Stanford Participatory Budgeting platform}{AnonPB Platform\footnote{The platform name has been anonymized for review}} is an online platform that has been used to engage residents in more than 150 budgeting processes. We present a data set with anonymized budget opinions from these processes with K-approval, K-ranking or knapsack primary ballots. For a subset of the voters, it includes paired votes with a different elicitation method in the same process. This presents a unique data set, as the voters, projects and setting are all related to real-world decisions that the voters have an actual interest in. With data from primary ballots we find that while ballot complexity (number of projects to choose from, number of projects to select and ballot length) is correlated with a higher median time spent by voters, it is not correlated with a higher abandonment rate. 

We use vote pairs with different voting methods to analyze the effect of voting methods on the cost of selected projects, more comprehensively than was previously possible. In most elections, voters selected significantly more expensive projects using K-approval than using knapsack, although we also find a small number of examples with a significant effect in the opposite direction. This effect happens at the aggregate level as well as for individual voters, and is influenced both by the implicit constraints of the voting method and the explicit constraints of the voting interface. Finally, we validate the use of K-ranking elicitation to offer a paper alternative for knapsack voting.
\end{abstract}

\section{Introduction}
Engaging residents/citizens in decision-making is a longstanding challenge \citep{langton_american_1979, ebdon_citizen_2006}, resulting in stakeholder influence ranging from merely being informed to gaining direct control \citep{arnstein_ladder_1969, fung_varieties_2006}. In the United States, citizen participation has been mandated and/or commonplace for many decades \citep{callahan_elements_2007, langton_american_1979} and many cities report some level of involvement by their citizens in the budgeting process \citep{wang_assessing_2001}. Citizen participation in budgeting processes has however been criticized as often lacking depth \citep{callahan_citizen_2007, van_dijk_digital_2012}. As society is embracing online technology, administrators are presented with an opportunity to better engage their residents. 

Participatory Budgeting (PB) stands out as an empowering mechanisms to engage residents and is globally increasingly adopted \citep{bartocci_journey_2022}. PB is associated with many different definitions, which have in common that they allocate a budget across budget items (projects) with stakeholder (citizen) participation in that decision \citep{williams_participatory_2017, miller_modes_2019, sintomer_participatory_2008}. In many cases in the United States, this will include a proposal, deliberation, and voting phase \citep{rudas_participatory_2021}. 

Theoretical research into vote aggregation depends on distribution assumptions, analytical work depends on the availability of clean and realistic data sets and practitioners depend on findings from both. With this paper, we publish a dataset that we believe can be valuable for research into different elicitation and aggregation methods. While such data could be collected from crowdworker platforms on mimicked budgeting elections, this would require assuming that real voters deciding on real issues would behave in the same way as people who have no real stake. 

We will focus our paper on the voting phase of Participatory Budgeting and especially focus on aspects that are relevant to PB organizers to design their ballot and make choices with regards to their voting method. We will specifically discuss some findings that we believe to be directly relevant to practitioners. While the data that we use in this paper is gathered on a fully functional platform, there is no reason why participatory budgeting can not be done via simple widgets on platforms such as X and Facebook. Some of the metrics that we study (such as completion time and completion rates) are perhaps even more relevant in these attention-poor social media settings.

\subsection{Definitions}
We will describe the setup of a voting process in a participatory budgeting \textit{election} on the \anonymize{Stanford Participatory Budgeting platform}{AnonPB Platform}. The election organizer determines the settings of the election including the voting method, available projects and their description, eligibility, and authentication. A voter $i \in \{1, ..., N\}$ is authenticated, and presented with a \textit{primary ballot} with $M$ projects $j$; typically defined by a title, description and cost $c_{j}$. 
The voter is invited to submit a \textit{vote} with approval values for each project $x_{ij}$, subject to constraints imposed by the \textit{elicitation method}. After submitting the primary ballot, a voter is sometimes given the possibility to submit a \textit{secondary ballot} which is used for research purposes only. Finally, they are often asked to submit a \textit{demographic survey} off-platform, which is not connected to their votes. 
The organizer can post-authenticate voters based on the authentication information they entered, and void any voters that are not meeting the criteria (e.g. residency in the city). An allocation is determined by the election organizer using an \textit{aggregation method} based on the remaining primary votes.

While the `voting method' is a single choice for the election organizer, it should be noted there are in fact three different choices combined: elicitation, consideration, and aggregation. Elicitation defines how votes are elicited from the voter, the consideration determines what information is exposed (and how prominently), and aggregation defines how the submitted votes result in an aggregate allocation. While consideration is a worthwhile endeavor to investigate in its own right, we will for simplicity consider it associated with the elicitation method. 

\begin{figure}[t]
\centering
    \includegraphics[width=0.7\linewidth]{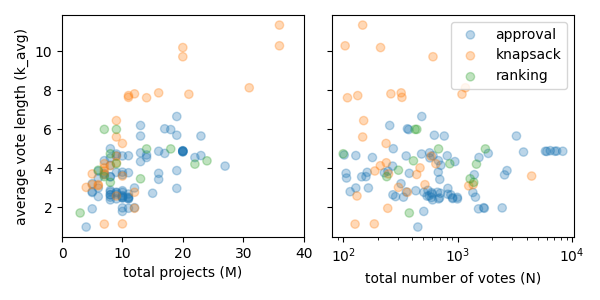}
    \caption{Elections in the dataset \\(1 election out of range)}
    \label{fig:pb elections m kavg}
\end{figure}
On this platform, three main elicitation methods are available with a default aggregation method associated with it. The available elicitation methods are:
\paragraph{K-approval voting}
Each voter selects (approves of) up to $K$ projects and submits a vote where $x_{ij} \in {0,1}$ and $\sum_j x_{ij} = k_i \leq K$. This is the traditional method used in American PB elections and 1-approval is identical to many regular elections in the United States. The advantage of this method is that it is intuitive for the voter. The main downside is that the voter can't easily trade-off projects of different cost: do they prefer one new playground of \$100,000 or rather upgrade two existing playgrounds for \$50,000 each?
\paragraph{K-ranking voting}
Each voter selects up to $K$ projects and ranks them in order of value for money. The advantage is that this is still intuitive, voters are increasingly familiar with ranked methods (the description is often simplified to `preference') and it allows voters to express more detailed preferences. The $K$ is often set to a (much) lower value than the number of projects: a full ranking is considered burdensome. 
\paragraph{Knapsack voting}
Each voter selects as many projects as they want, as long as the total cost of their selection does not exceed an available budget $B$ (fits in their ‘knapsack’). Their ballot is constrained by $\sum_j c_{j}x_{ij} \leq B$.
An advantage of this method is that this is the most natural way for voters to make trade-offs between projects and consider their cost. They make the same choice as a policymaker would. Online, this is an easy task, but on paper we can not implement this the same way. We will later discuss how we could indirectly elicit a knapsack vote through K-ranking, which is easier on paper.

The platform also supports \textbf{pairwise comparison voting} (a sequence of project pairs is presented to the voter, and they are asked to indicate their preference between the two) and \textbf{K-token voting} (each voter gets $K$ tokens to be distributed among the projects; an implementation of \textit{cumulative voting}) but these methods are primarily research methods and currently not actively used in elections. We will not consider these methods in this overview. 

Votes have to be aggregated to result in an allocation, and we will describe how this is implemented on the platform. For K-approval and knapsack elicited ballots, if the projects can only be selected integrally, the projects are ranked in decreasing order of received votes. The budget is distributed in that order. If projects can also be approved partially by a voter, projects are split into single-dollar projects and assigned one point for each voter that approved of them. 

Ties can be resolved arbitrarily, and only matter if they concern the last project to be allocated when the budget runs out. How remaining funds are distributed, depends on the context of the election organizer. For example, we can assume a total budget of \$100 and three projects in an aggregate ranking: A (\$70), B (\$50), C (\$30). Partial allocation to the next project is theoretically straight forward and in our example we would allocate \$70 to project A and \$30 to project B. If partial funding is not an option, the organizer could choose to continue down the ranking until a project is found that can be fully funded with the remaining funds (\$70 for project A and \$30 for project C) or the organizer could decide to not spend the remaining funds, but use those to, for example, increase the budget for the next cycle (\$70 for project A and \$30 remaining funds). 

For votes with ranking elicitation, it is less obvious what the appropriate aggregation method would be. A few families of aggregation methods for ranked ballots are discussed in this paper:
\begin{itemize}
    \item \textbf{Inferred K-approval}: Assign score 1 to all projects that receive up to rank K, and score 0 to all other projects. 
    \item \textbf{Borda count}: Assign score $M - k$ to a project with rank $k$ (with $0.5(M-K_i-1)$ score for unranked projects, where $K_i$ is the number of ranked projects by voter $i$) \citep{de_borda_memoire_1784}. We also consider some variations on the traditional Borda count. \textit{Borda-MK} does not assign any score to unranked projects, assuming that voters don't have value for them. \textit{Borda-K1} assigns score $K_i$ to the top-ranked project, score 1 to the last-ranked project and no score for unranked projects. 
    For each of these variations, we can replace $K_i$ with $K$, assuming that each voter fills out their complete ballot. The aggregation code for our platform uses a full Borda count with $K$ at the election level. 
    \item \textbf{Inferred Knapsack}: Sequentially assign budget to a project in order of the rank for that voter until the budget is exhausted \citep{Goel2019}. Either use the remaining budget to partially fund the next project (ranking-partial), or skip a project that cannot be funded in favor of a possibly less expensive project that is ranked lower by the voter (ranking-skip). This choice depends on the assumption whether projects can be partially funded or not. 
\end{itemize}

This will then result in an aggregate list of the projects in decreasing order of received votes and budgets are then assigned as above.

\section{Related work}
Participatory Budgeting has a rich literature with case studies and evaluations both by academics and practitioners, offering insights into its potential and effects (e.g. \citep{su_porto_2017, falanga_participatory_2020}). This literature is primarily discussed in two fields: in the Public Administration and Political Sciences, PB is approached as a democratic innovation or administrative process to engage stakeholders (citizens/residents), encompassing idea-collection, project shaping, voting and implementation \citep{bartocci_journey_2022, hagelskamp_public_2016}. In the social choice literature, PB serves as an important use case for aggregating budgeting opinions \citep{rudas_participatory_2021, rey_computational_2023}. 

PB spread across the globe as a democratic innovation since it was introduced in 1988 in Porto Alegre, in many different flavors \citep{sintomer_participatory_2008}, with a comprehensive survey of its history in \citep{bartocci_journey_2022}. For North America, where most of our data originates, a comparative study of PB evaluations is available in \citep{hagelskamp_public_2016}. 
The goals of PB vary widely and in the US context they can explicitly include equity \citep{lerner_budgeting_2020}. Engagement with PB has been found to increase political participation of traditionally underrepresented demographics in New York \citep{johnson_testing_2021} and engagement in budget feedback exercises can enhance understanding of budgetary issues \citep{kim_budgetmap_2016}. 
As PB increasingly moves online, various tools including the \anonymize{\anonymize{Stanford Participatory Budgeting platform\footnote{https://pbstanford.org}}{AnonPB Platform\footnote{Anonymized URL}} \anonymize{\citep{Goel2019}}{(Anon et al., 20xx)}, Consul\footnote{https://consulproject.org} \citep{arana-catania_citizen_2021, pina_decide_2022}, AppCivist-PB\footnote{https://pb.appcivist.org} \citep{holston_engineering_2016} and Decidim \citep{serramia_optimising_2019}}{(4 Anonymous platforms including AnonPB with citation and link)} become available. It would go beyond the scope of this paper to attempt to make a complete inventory or make an extensive comparison, but some tools have been compared in \citep{the_democratic_society_digital_2016, cantador_towards_2018}.

Knapsack voting and PB have gained interest in the algorithms and social choice community over the last decade, with PB as an extension to (continuous or discrete) multi-winner elections. Detailed surveys into voting methods are available in \citep{rey_computational_2023, rudas_participatory_2021}. Beyond voting methods, there has been an interest in how votes can be aggregated once elicited, for example through a method that mirrors more deliberative approaches \citep{goyal_lowsample_2023}, considering fairness measures such as the core \citep{fain_core_2016, munagala_auditing_2022}, and justified representation such as the Method of Equal Shares \citep{peters_proportional_2021}.

Vote elicitation in a budgeting context has been explored with theoretical, experimental, and data-driven approaches. 
In \citep{benade_preference_2017} 4 elicitation methods are discussed: K-approval, K-ranking (in two variants), and knapsack voting, concluding among others that knapsack voting might be too burdensome for voters and does not have as many theoretical guarantees. However, \citep{Goel2019} suggests that knapsack voting was less burdensome in practice and desirables strategic properties. 
In \citep{garg_who_2019} learning rates were proposed to identify optimal ballot design, concluding that the threshold $K$ was historically set too low in $K$-approval voting for optimal learning. A study into the effect of K-approval, threshold-approval, K-ranking (in two variations), knapsack and K-token voting with recruited crowdworkers evaluated the effect on cognitive load and voters' ability to recall their stated preferences \citep{fairstein_participatory_2023}, suggesting Value-for-Money ranking as cognitively the hardest on voters. In contrast to their work, our findings are based on real world PB processes.
 
Unfortunately, voter preference data from PB elections is scarce, and empirical work spanning multiple elections is limited. A notable exception is the Pabulib, an open library with vote sets from over 800 PB elections, mostly from Poland \citep{stolicki_pabulib_2020} with mostly data from $K$-approval ballots and $K$-ranking ballots with low $K$. We believe that our data will be a valuable addition, as we report the time spent on the ballot by the voter, have different voting methods available and have votes available from secondary voting methods. This provides both vote pairs (two votes from the same voter on the same projects with different voting methods) and ballot pairs (where voters were randomly assigned to a voting method for their secondary ballot). 

\section{Platform and data collection}
The \anonymize{Stanford Participatory Budgeting platform}{AnonPB Platform} has been previously described in \anonymize{\citep{Goel2019}}{(Anon et al., 20xx)} and is one of several platforms that cities in North America have available to organize an online voting phase in their PB election. 
The software is available under a free license on GitHub and could be installed by an election organizer on their own servers. One instance of the platform can support many ballots, with admin permissions at the election level. This way the \anonymize{Crowdsourced Democracy Team}{Anonymous Group} has been able to provide the system to multiple organizers, giving each the possibility to manage their own election and configure their ballot.

Our data set only contains data from elections that were hosted on the instance that is provided by the \anonymize{Crowdsourced Democracy Team}{Anonymous Group}. A ballot instance is provided at no cost, and the organizer can then set up the ballot in their desired configuration. Some of the available configuration options are:
\begin{itemize}
    \item Voting method: K-approval, K-ranking, knapsack or K-token
    \item Language: Currently, the ballot can be made available in: English, Amharic, Arabic, Bengali, Chinese, Finnish, French, German, Haitian Creole, Hindi, Hmong Daw, Khmer, Polish, Portuguese, Spanish, Tagalog. Select a subset of the languages, with one language as default. The voter can then vote in any of the selected languages.
    \item Voter validation: SMS confirmation, personal information, generated codes, free-form (no verification). 
    \item Voter registration: Questions that the voter has to answer and that can be used for post-voting verification
    \item Voting phases: voting phases that can be presented
    \item Available budget
    \item Voting constraints (value of $K$), as appropriate for the voting method
    \item Project appearance: whether to show the cost, project numbers and maps
    \item Survey link for demographic survey off-platform
\end{itemize}

An election is always configured with a primary voting method (the official ballot) and the authors have in some elections (when approved by the organizer) also set up a secondary voting method (the research ballot).
% \footnote{The K-tokens voting method is still experimental and has not been used as primary voting method.} 
If a secondary voting method is set up, the voter is first presented with the primary ballot, before being presented with a consent text and a secondary voting method (with the same projects). The secondary method can be set up as an array of voting methods, in which case the voter is randomly presented with one of the voting methods in the array. 

The organizer also configures the content of several city-specific fields, including the landing page text, the contact information, in-person voting date, and location. The organizer also enters the voting dates, projects (title, description, cost (discrete or with cost steps), and optionally location, category, image, video URL, project details, and coordinates. 

\section{Data}
\label{sec:data}
% \begin{figure}[t]
%     \includegraphics[width=0.95\linewidth]{figs/elections_n_kavg.png}
%     \caption{Elections in the dataset \\(1 out of range)}
%     \label{fig:pb elections n kavg}
% \end{figure}
We present a data set of anonymized ballots from the \anonymize{Stanford Participatory Budgeting platform}{AnonPB Platform} described above.\footnote{\anonymize{The data set is available through the Stanford Digital Repository: https://doi.org/10.25740/db709zg9088 \citep{gelauff_participatory_2024}}{The data set is available through the Anonymous Repository: anonymous url}} Since 2014 we have partnered on 150+ ballots with 
% TODO with YYY organizations, 
mostly local governments and foundations in the United States, who used the platform to organize a Participatory Budgeting election. The organizer (partner organization) could configure their election ballot to their own specifications, with their own projects and in the languages that they desired. The organizer was also in charge of recruitment among their stakeholder population and implementing the results. 

From the voter's perspective, they arrive on a landing page, authenticate, access the primary ballot, and optionally access a secondary ballot and/or a survey. The voter could exit at any point and may have the option to skip a phase. We do not have access to the information from the demographic survey that was provided by the city and would not be able to connect this back to the voting data. 

For this data set, we provide tables \textit{elections} (and a version enriched with summary statistics), \textit{projects}, \textit{voters}, \textit{vote\_approvals}, \textit{vote\_knapsacks}, \textit{vote\_rankings}, and \textit{vote\_tokens} as of 22 May 2023. We have added some additional columns and tables with derived data for easier analysis. Elections, voters, projects and votes have been cleaned to ensure quality, and data was removed when that was necessary for anonymization. We describe and justify these steps in the data set documentation. 

We present data from 124 primary ballots (79 approval, 32 knapsack, 13 ranking) and 38 secondary ballots (2 approval, 18 knapsack, and 18 ranking)
% update 240109 stats_data_set_count
with at least 100 votes per election. Some elections have data from two secondary ballots. Aggregations of this data may be different from published results, due to more aggressive data cleaning for privacy and data quality purposes, as the purpose of this data set is to understand voting behavior. 
The following tables are included:
\begin{itemize}
    \item elections: election ID, election name and available budget, text on the ballot, remarks
    \item elections\_rich: the same as elections, but with additional details about ballot settings and summary statistics
    \item projects: project ID, election ID, category ID, ordering, cost and geo coordinates (when available)
    \item voters: voter ID, election ID, authentication method, last stage entered, last day updated
    \item votes: voter ID, election ID, project ID, allocated budget, rank (ranking only), tokens (token only), last day updated, fraction of voters that completed the vote in less time
    \item inferred votes: inferred knapsack votes when a ranking vote would be used to fill out a knapsack ballot. 
    \item voter utility stats: statistics describing the utility overlap between vote pairs
\end{itemize}
The voter ID has been altered and the timestamp information is rounded or removed for anonymization purposes. In the data set documentation (``the documentation'') we describe the tables and the data cleaning process in more detail.
% update 230715

% In Table~\ref{app:tab:pb dataset elections} in the Appendix we describe the elections that are available in the dataset. In Tables~\ref{app:tab:pb dataset approval}, \ref{app:tab:pb dataset knapsack} and~\ref{app:tab:pb dataset ranking} we provide the number of ballots per voting method (including both primary and secondary voting method). 

% TODO: add response rates? Dropoff rates?
% \subsection{Descriptive analysis}
% <rough editing>
% * What can we tell about the elections by just looking at them?* How many voters per election. How does that compare to the number of people estimated to be eligible voters?
% * How many projects do they typically put on the ballot? What does the budget distribution across these projects look like?
% * How did the selected elicitation method evolve over time?

In Figure~\ref{fig:pb elections m kavg} we visualize the primary ballots in our dataset, with the number of projects that voters were able to choose from (M), the number of voters that cast a vote (N) and the average number of selected projects on that ballot ($k_{avg}$). 

The data set may prove useful as a resource for realistic voting data under different budget-related voting methods and to understand vote distributions under different conditions. Unique aspects of this data are the relative amount of time spent (as quantile of the population) and the fact that vote pairs under different voting methods from the same voter on the same election are present. Users of the data should be aware that the voters usually only represent a small portion of the eligible population, although the exact size of this portion has not been estimated by the authors. Recruitment methods and other factors can therefore have a significant effect on composition of the set of voters \citep{gelauff_advertising_2020}.

% TODO: describe some statistics on the ballot design.
% * How many projects m? What is k? 
% * How long are the titles? Descriptions?
% * How often are there images?
% * How often is the project discrete vs partially fundable?
\section{Comparing vote pairs}
\label{sec:pb analysis}
We will establish that secondary ballots meaningfully describe the opinion that was expressed by the same voter on the primary ballot. This makes the comparison of vote pairs possible, but also allows us to consider secondary ballots in the first place, which is especially helpful as they contain $K$-ranking ballots with a relatively high $K$.
Next, we will establish that K-approval and knapsack ballots can be inferred from sufficiently long K-ranking ballots. 

We compare the primary and secondary votes to better understand the voting behavior with different voting methods at a granular level. We use inferred knapsack and $K$-approval votes to compare different aggregation methods, and how their constraints affect the individual choices of a voter and the eventual allocation after aggregation. 

\subsection{Secondary ballots}
\label{sec:secondary ballots}
% update 230715 df_describe_util, df_describe_util_kn
\begin{table*}[t!]
\small
\begin{tabular}{p{0.045\textwidth}p{0.045\textwidth}p{0.05\textwidth}|p{0.032\textwidth}p{0.038\textwidth}p{0.04\textwidth}|p{0.032\textwidth}p{0.038\textwidth}p{0.04\textwidth}|p{0.032\textwidth}p{0.038\textwidth}p{0.04\textwidth}|p{0.032\textwidth}p{0.038\textwidth}p{0.04\textwidth}}
\toprule
Prim. & Sec. & n & \multicolumn{3}{p{0.12\textwidth}|}{per-election median percentile} & \multicolumn{3}{p{0.11\textwidth}|}{per-voter percentile} & \multicolumn{3}{p{0.11\textwidth}|}{per-election median z-score} & \multicolumn{3}{p{0.11\textwidth}}{per-voter z-score} \\
& & & med & mean & min & med & mean & std & med & mean & min & med & mean & std  \\ \midrule
app & knap & 16 & 1.00 & 0.99 & 0.90 & 1.00 & 0.92 & 0.17 & 1.37 & 1.44 & 1.07 & 1.41 & 1.44 & 0.99 \\
app & rank & 11 & 1.00 & 1.00 & 0.98 & 1.00 & 0.93 & 0.17 & 1.11 & 1.25 & 0.80 & 1.21 & 1.16 & 0.80 \\
app & tok & 1 & 1.00 & 1.00 & 1.00 & 1.00 & 0.93 & 0.16 & 1.25 & 1.25 & 1.25 & 1.25 & 1.00 & 0.69 \\
knap & app & 1 & 0.88 & 0.88 & 0.88 & 0.88 & 0.76 & 0.27 & 1.12 & 1.12 & 1.12 & 1.12 & 0.93 & 1.15 \\
knap & rank & 7 & 1.00 & 0.97 & 0.89 & 0.99 & 0.82 & 0.25 & 1.06 & 1.11 & 0.68 & 0.95 & 0.88 & 0.84 \\
% rank & app & 0 & NaN & NaN & NaN & NaN & NaN & NaN & NaN & NaN & NaN & NaN & NaN & NaN \\
rank & knap & 1 & 1.00 & 1.00 & 1.00 & 1.00 & 0.92 & 0.16 & 2.25 & 2.25 & 2.25 & 2.25 & 2.11 & 1.25 \\
all & all & 26 & 1.00 & 0.99 & 0.88 & 1.00 & 0.91 & 0.19 & 1.26 & 1.30 & 0.68 & 1.28 & 1.31 & 0.98 \\
\midrule
knap & i-kn$_p$ & 7 & 0.89 & 0.91 & 0.81 & 0.92 & 0.82 & 0.24 & 1.28 & 1.26 & 0.93 & 1.27 & 1.09 & 1.02 \\
knap & i-kn$_s$ & 7 & 0.97 & 0.96 & 0.88 & 0.96 & 0.82 & 0.24 & 1.30 & 1.39 & 1.22 & 1.36 & 1.11 & 1.08 \\
i-kn$_p$ &knap & 1 & 0.99 & 0.99 & 0.99 & 0.99 & 0.92 & 0.16 & 2.27 & 2.27 & 2.27 & 2.27 & 2.12 & 1.26 \\
i-kn$_s$ &knap & 1 & 0.99 & 0.99 & 0.99 & 0.99 & 0.92 & 0.16 & 2.26 & 2.26 & 2.26 & 2.26 & 2.09 & 1.28 \\
\bottomrule
\end{tabular}
\caption{Utility overlap statistics of primary (prim) / secondary (sec) ballot pairs: K-approval (app), knapsack (knap), K-token (tok), K-ranking (rank), inferred knapsack with partial (i-kn$_p$) and skip (i-kn$_s$) allocation. Per-election statistics report the median, mean, and minimum of the per-election median utility overlaps. Per-voter statistics give the median, mean, and standard deviation of all individual utility overlaps. n is the number of elections for that ballot pair.\\ 
% Percentiles are inclusive, and z-scores indicate the median z-score (standard deviations from the mean utility overlap) for a voter in the context of other ballot pairs in the same election.\\
% The same election can have multiple ballot pairs; 
}
\label{tab:pb utility overlap stats}
\end{table*}
We evaluate how similar the primary ballot of voter $i$ ($x_i$) is to their secondary ballot ($y_i$) compared to the secondary ballot of all voters $i'$ ($y_{i'}$). The budget overlap utility of a voter is the overlap between a vote and some other set of projects (e.g. the final allocation) in dollars. We calculate the budget overlap utility between any possible vote pair (irrespective of the voting method used) by finding for each project the minimum cost approved between the two votes, and taking the sum. We rank the utility $u_{ii}$ among all $u_{ii'}$ and find its percentile (inclusively: proportion of utilities that are equal or lower) and z-score (the number of standard deviations that $u_{ii}$ is from the average $u_{ii'}$). We calculate for each election the median percentile and median z-score, and report for each combination of voting methods the median, mean, and minimum across all per-election scores as well as the median, mean, and standard deviation per-voter across elections. Finally, we do the same for pairs of knapsack votes and inferred knapsack votes from K-ranking votes (Table~\ref{tab:pb utility overlap stats}).
% \footnote{The fact that the number of ballot combinations does not add up, is that in some elections, voters were randomly assigned between to secondary ballots} 

We observe that the overlap percentile, as well as per-election median percentile, is in most elections 1 (perfect overlap between primary and secondary ballots). For each combination, the per-election median percentile is 0.88 or higher. We also observe that the average percentile is 0.88 or higher for each ballot combination. 
% One could however argue that a high percentile is no guarantee of a good match: it could just mean that everyone submits an identical set of projects in their secondary ballot. 
% This is why we also look at the z-score. 
We observe that the median z-score is in most elections 1 or higher (where a positive z-score means higher than average overlap), and the mean z-score is around 1 for each voting method combination. 

Because these are different voting methods, we would expect some differences to appear between different ballots of the same person -- even when the voters have perfect memory and attention span. These numbers indicate that there is a signal in the secondary ballots and that they are not randomly entered, and in subsequent analysis we found no counter indications that there is a substantive difference between primary and secondary votes. Going forward, we will for brevity sake consider primary and secondary ballots equally, unless otherwise mentioned. 

\subsection{Different aggregation methods for K-ranking ballots}
\label{sec:pb aggregating}

\begin{table*}[t!]
% update 240109 cosine
\small
\centering
    \includegraphics[width=0.7\linewidth]{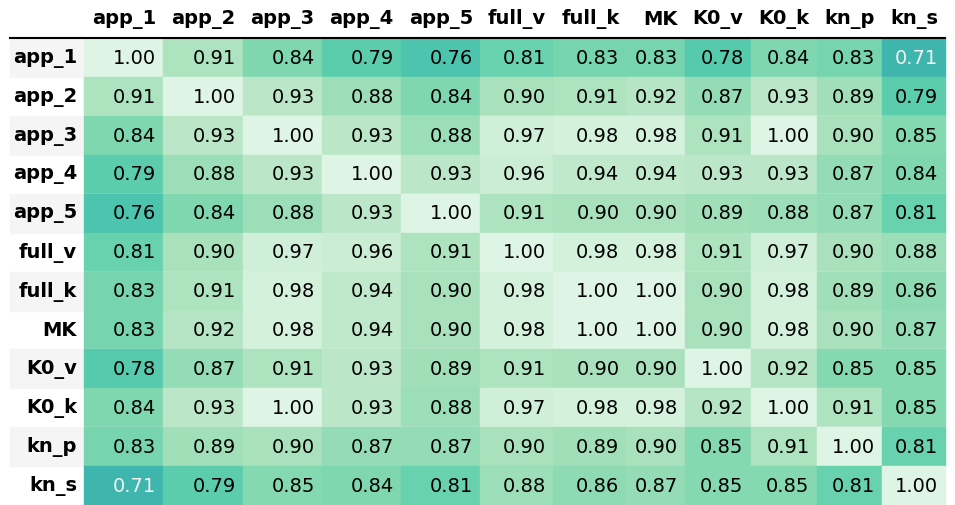}
\caption{Cosine similarity between vectors of projects with the proportion of the budget allocated under different aggregation methods, inferred from identical K-ranking ballots: K-approval (app$_K$), full Borda ranking, Borda-MK, Borda-K1 (with voter- ($_v$) and election-specific $K$'s ($_k$)), knapsack with partial (kn$_p$) and skip-allocation (kn$_s$).  Colors indicate magnitude.}
\label{tab:pb cosine ranking aggregations}
\end{table*}

In our dataset, we have 31 elections 
% update 240109: number_ranking_ballots
with ranking ballots. With some natural assumptions we can use ranking votes to infer how voters would have responded to different voting methods, if they would have followed that ranking. This allows us to explore the effect of the implicit constraints that a voting method imposes on a voter, as this approach excludes any effect from the information available on the ballot or strategic behavior. 

For K-approval, we make the assumption that if the voter had been asked to select $K' \leq K$ projects, they would have submitted the $K'$ highest-ranked projects on their K-ranking ballot. For knapsack aggregation, we assume that the voter would have gone down their ranking, adding projects one at a time as the budget constraint permitted. We extract this way what we will call inferred approval and knapsack votes, and then aggregate them as described above for K-approval and knapsack aggregation. While this is a useful approximation, this assumption might not always hold: a voter might for example want to consider interaction effects between projects, or the available ranking might not be long enough. 

For each of these elections with ranking ballots, we aggregate the ballots with K-approval ($K$ = 1, ..., 5), 5 variations of Borda count, knapsack-partial and knapsack-skip aggregation methods. For each aggregation method, we create a vector of the portion of the available budget that was allocated to each project with that aggregation method and calculate the cosine similarity between these vectors (see Table~\ref{tab:pb cosine ranking aggregations}). Cosine similarity scores range between 0 and 1, where a score of 0 indicates that the vectors are orthogonal, and the larger the score is, the smaller the angle between the vectors.

This is a crude comparison but effectively shows that across all available elections, the aggregation differences between the different members of the Borda family are small with a cosine similarity of 0.92 or higher. We observe that the allocations can be rather different for different choices of $K$ under K-approval voting.  

\subsection{Knapsack aggregation from ranked ballots}
\label{sec:pb knapsack aggregation}
Knapsack voting is particularly hard to achieve on paper, where it would be unrealistic to expect voters to constantly calculate whether their selection of projects fits within the allowed budget. In previous work \citep{Goel2019} it was argued that organizers could effectively provide a paper knapsack ballot, by eliciting a ranked ballot with a large $K$ and infer the knapsack ballot for the voters. The cognitive load that input formats would place on voters has been raised as a concern and it was suggested that experiments would be needed \citep{benade_preference_2017}. With our data we can attempt to answer this concern.

We have 8 elections with a knapsack and K-ranking ballot pair for the same voters.
% We have access to 7 elections with a knapsack primary ballot and a K-ranking secondary ballot, and 1 election with a K-ranking primary ballot and a knapsack secondary ballot where both primary and secondary ballot have sufficient votes available. 
We calculate the utility overlap between these votes and report the same statistics as before (Table~\ref{tab:pb utility overlap stats}). 
We observe that the percentile scores for utility overlap between the knapsack ballot and the inferred knapsack ballot are high, and that using the `skip' method results in inferred votes closer to the knapsack votes. This is expected, as this is in line with what a voter could do in practice (partial allocation is often not allowed) and going forward we will use the `skip' method unless otherwise specified. The z-scores are on average higher than 1, indicating that inferring a knapsack ballot from a ranking ballot is indeed a valid way of achieving a knapsack ballot when online methods are not available. However, while this is a good substitute, it is not identical: we will discuss below the effect that the interface (knapsack compared to ranking) can have on the selected projects.

% \subsubsection{Similarity between the outcomes from different aggregation methods }
% \lodewijk{this needs more description}
% When we aggregate all the elections with a large ranking available using the various aggregation methods, we can compare the various aggregation methods that are possible. 

% \subsubsection{Inferring approval ballots}
% \lodewijk{I'm removing this subsection, as the conclusions are unclear and this is probably not the best approach to this question. Future work.}
% Similarly as \citep{garg_who_2019} we can now infer several different approval ballots from the provided rankings. With this information, we can see what the impact is of different choices of 'k' in approval voting. 
% * Do people make different choices if they are asked for a lower or higher k? TODO

\section{Project cost}
\label{sec:pb project cost}
Voters may consider project costs differently when confronted with explicit budget constraints. Goel et al. \citep{Goel2019} found that K-approval voting favored more expensive projects in aggregate vote distributions, resulting in a higher average cost for winning projects compared to knapsack voting. This aligns with the notion that voters would be more frugal when they encounter explicit budget constraints. This analysis focused on aggregate data and did not distinguish between explicit constraints considered by the voter, and implicit constraints imposed by the voting method. Our data allows for a more nuanced exploration.

We aim to dissect the effect of the voting interface and of the voting method. Using elections with ballot pairs, we compare the cost of selected projects using votes under different voting methods by the same voter. We isolate the effect of the user interface and provided information on the ballot by comparing knapsack votes with inferred knapsack votes from the same voter. Similarly, we isolate the impact of the voting method itself by inferring votes under different voting methods from the same K-ranking ballots. 

We report our cost statistics consistently per ballot pair (e.g. knapsack and inferred knapsack). We calculate the cost statistics (most expensive selected project, average cost of the 3 most expensive selected projects and average cost of all selected projects) for each vote and report for each ballot pair per election the average difference of cost statistics, normalized by the budget available in that election. Significance is established through a percentile-bootstrap (1000 simulations) per election and a two-sided test at 95\% (*) and 99\% (**) confidence intervals, recognizing that the statistics may not follow a normal distribution. 

To understand whether this effect is dominated by a small number of voters that made large shifts, we calculated for each election how many voters had a higher cost statistic for the first (K-approval) than for the second method (knapsack) and counted the number of elections where these voters formed the majority. We did this for each ballot pair comparison in this paper, and reported the numbers in Table~\ref{tab:pb cost net shift overview}. 

\subsection{K-approval and knapsack voting}
\label{sec:pb approval knapsack}
\begin{table}[t!]
% update 240109 df_costshift_app_kn
\small
    \centering
\begin{tabular}{lrrrrl}
\toprule
    {id} & n & top-1 & top-3 & avg & Prim. \\
\midrule
4 & 146 & 0.004 & -0.016 ** & 0.001 & app \\
5 & 141 & -0.006 & -0.004 & 0.016 ** & app \\
14 & 162 & 0.009 & 0.004 & 0.011 * & app \\
17 & 940 & 0.024 ** & 0.039 ** & 0.029 ** & app \\
103 & 231 & 0.013 & 0.067 ** & 0.097 ** & app \\
119 & 175 & 0.033 ** & 0.103 ** & 0.132 ** & app \\
121 & 149 & 0.005 & 0.064 ** & 0.073 ** & app \\
122 & 220 & 0.007 & 0.098 ** & 0.141 ** & app \\
126 & 280 & -0.029 * & -0.0 & 0.035 ** & app \\
128 & 234 & -0.029 ** & -0.027 ** & -0.018 * & app \\
130 & 517 & 0.01 ** & 0.006 ** & 0.01 ** & app \\
152 & 204 & 0.09 ** & 0.062 ** & 0.116 ** & knap \\
170 & 267 & -0.051 ** & -0.034 ** & -0.003 & app \\
172 & 405 & -0.006 & -0.016 ** & 0.033 ** & app \\
173 & 246 & -0.038 ** & -0.023 ** & -0.018 ** & app \\
174 & 447 & -0.006 ** & 0.005 ** & 0.005 ** & app \\
255 & 472 & 0.063 ** & 0.033 ** & 0.033 ** & app \\
\midrule
avg & 17 & 0.004 & 0.021 & 0.041 & \\
\bottomrule
\end{tabular}
    \caption{Average cost difference between pairs of approval and knapsack votes from the same voter. A positive score implies that the statistic for K-approval was that portion of the election budget higher. The bottom row shows the election count and average election statistics.}
    \label{tab:pb cost shift approval knapsack}
\end{table}

\begin{table}[t]
    \centering
    \small
    \begin{tabular}{p{0.035\textwidth}p{0.03\textwidth}r|rrr|rrr}
    \toprule
    ballot~1 & ballot~2 & n & \multicolumn{3}{c|}{ballot 1 $>$ ballot 2} & \multicolumn{3}{c}{ballot 1 $<$ ballot 2} \\
    & & &top-1 \!\!\!\!\!\!&top-3 \!\!\!\!& avg \!\!& top-1 \!\!\!\!\!\!& top-3 \!\!\!\!& avg \!\!\\ 
    \midrule
        app & knap      & 17    & 10     & 9     & 16    & 7 & 8     & 1 \\ % Approval and Knapsack
        knap & \mbox{i-kn$_{s}$} & 8       & 4     & 7     & 1     & 2 & 1     & 7 \\ % Knapsack and inferred knapsack
        app & \mbox{i-app} & 11        & 10    & 8     & 6     & 0 & 1     & 3 \\ % Approval and inferred approval
        \mbox{i-app$_4$} & \mbox{i-kn$_s$} & 31    & 19    & 15    & 23    & 7 & 11    & 2 \\ % inferred 4-approval and inferred knapsack
    \bottomrule
    \end{tabular}
    \caption{Per ballot pair, the number of elections with more voters that had a higher average cost statistic for method 1 than for method 2, and vice versa. Inferred ballots are indicated with {`i-'}.}
    \label{tab:pb cost net shift overview}
\end{table}

We consider 17 elections with sufficient voters who submitted both an approval and a knapsack vote and consider the cost statistics for their selected projects (see Table~\ref{tab:pb cost shift approval knapsack}). Except for one election, the primary voting method was K-approval and the secondary method was knapsack. 

In our per-voter analysis we validate Goel et al.'s finding: in most elections, the K-approval ballot resulted in a significantly higher average cost than the knapsack method. However, there are two instances where the knapsack ballot has a significantly higher average cost for selected projects.
This dominant effect can be explained by the explicit constraints presented in the knapsack voting interface. As voters select their preferred projects, they will have some budget left over, which they can only allocate to a low-cost project. Under K-approval voting, no such pressure exists. This pressure is however not unnatural: it mirrors constraints faced by decision-makers, and can even be argued to be desirable. 

This effect is much smaller when we focus on more expensive projects that the voter selected (top-1 and top-3 average costs). In several elections the shift even reverses direction: the voter selected a more expensive project with knapsack than with K-approval. This suggests that the magnitude and direction of this effect depend on the specifics of the election, especially concerning more expensive projects. This trend is even more pronounced in Table~\ref{tab:pb cost net shift overview}: in almost every election, voters with more expensive projects on their K-approval ballot were in the majority -- but this is not true when we only look at their most expensive selected projects.

These observations align with the explanation that voters are compelled to be more budget-conscious due to budget constraints after selecting their most expensive preferences. If they would choose only one instead of two of the most expensive projects under knapsack constraints, we would expect to see a strong effect in the top-3 statistics (the average number of selected projects $K_{i}$ is usually below 5). 

% For instance, if there are two expensive projects to choose from, they may only choose one on the knapsack ballot, and both on the K-approval ballot. 
% if there are two expensive projects to choose from, they may only choose one on the knapsack ballot, and both on the K-approval ballot. However, this explanation no longer holds when we consider the effect on the top-3 most expensive projects that the voter selected. If this were indeed the case, we would expect a similar number of voters to shift in that statistic given that the average number of projects $k_i$ is for most of these elections below 5. This dynamic is very different depending on the exact projects on the ballot, and therefore it makes sense that there are significantly different costs for the most-expensive project on a voters' ballot in both directions, depending on the exact available projects.

\subsection{Effect of the voting interface}
\label{sec:pb voting interface}
% updated 240109 df_costshift_kn_knsim

\begin{table}[t]
\small
    \centering
    \begin{tabular}{lrrrrl}
    \toprule
    {id} & n & top-1 & top-3 & avg & Prim. \\
    \midrule
93 & 740 & 0.003 ** & 0.004 ** & -0.014 ** & rank \\
152 & 226 & -0.012 & -0.005 & -0.09 ** & knap \\
171 & 91 & 0.011 & 0.012 & 0.016 & knap \\
194 & 137 & 0.007 & -0.001 & -0.019 ** & knap \\
248 & 106 & -0.002 & 0.0 & -0.015 & knap \\
249 & 352 & 0.005 & 0.005 & -0.006 * & knap \\
250 & 408 & 0.006 & -0.006 & -0.022 ** & knap \\
251 & 127 & 0.013 & 0.013 * & -0.014 * & knap \\
\midrule
avg & 8 & 0.002 & 0.002 & -0.020 & \\
    \bottomrule
    \end{tabular}
    \caption{Average cost difference between pairs of knapsack votes and inferred knapsack (skip) votes. A positive score implies that the statistic for knapsack was higher.}
    \label{tab:pb cost shift knapsack inferred}
\end{table}

% \begin{table}[ht]
% \small
%     \centering
% \begin{tabular}{lrrrr}
% \toprule
%     {id} & n & top-1 & top-3 & avg \\
% \midrule
% 93          &  606 &    0.02 &    0.06 &  -0.39 \\
% 152         &  151 &   -0.03 &    0.21 &  -0.58 \\
% 171         &   75 &   -0.01 &    0.05 &   0.09 \\
% 194         &   96 &    0.03 &    0.12 &  -0.33 \\
% 248         &   77 &   -0.12 &   -0.04 &  -0.18 \\
% 249         &  243 &    0.02 &    0.05 &  -0.10 \\
% 250         &  276 &    0.00 &    0.12 &  -0.25 \\
% 251         &   86 &    0.03 &    0.13 &  -0.14 \\
% \midrule
% $> 0$ & 8 & 4 & 7 & 1 \\
% $< 0$ & 8 & 3 & 1 & 7 \\
% \bottomrule
% \end{tabular}
%     \caption{Net portion of voters who had a higher average cost statistic for their knapsack ballot than for their inferred knapsack ballot. A negative number implies that there was a net portion where the average cost was higher on their inferred knapsack ballot. The table describes the average project cost of the top-1, top-3 and all submitted projects on the ballot.}
%     \label{tab:pb cost net shift knapsack inferred}
% \end{table}

In order to isolate the effect of the voting interface and information provided to the voter, we compare knapsack votes (from the knapsack interface) with inferred knapsack votes (from the K-ranking interface). This comparison excludes any effects from implicit constraints posed by the voting method. We report the cost statistics in Tables~\ref{tab:pb cost shift knapsack inferred} and \ref{tab:pb cost net shift overview}. 

If explicit constraints and the information provided to the voter were responsible for the observed cost differences between K-approval and knapsack, we would expect a similar shift here. Indeed, we observe a consistent trend for the overall average cost: the average cost is higher for inferred knapsack votes than for knapsack votes. However, the top-3 statistics in Table~\ref{tab:pb cost net shift overview} give a more nuanced picture: almost all elections had a majority of voters select more expensive top-3 projects with explicit knapsack constraints than without. This effect is opposite of what we would expect.

This suggests that while voters may exhibit greater cost-consciousness when confronted with explicit budget constraints, this effect primarily manifests among less expensive projects on the ballot. Among more expensive projects, we do not observe significantly different average project costs in most elections. In cases where the average top-3 project cost was higher on the knapsack ballot than on the inferred knapsack ballot, it could be hypothesized that voters prefer to 'upgrade' more expensive projects while 'downgrading' or even splitting less expensive ones.

% updated 240109 df_costshift_app_appsim
\begin{table}[t]
\small
    \centering
    \begin{tabular}{lrrrrl}
    \toprule
    {id} & n & top-1 & top-3 & avg & Prim. \\
    \midrule
103 & 247 & 0.016 & 0.014 & 0.014 & app \\
118 & 118 & 0.026 ** & 0.015 ** & 0.015 ** & app \\
119 & 152 & 0.011 & 0.017 * & 0.017 * & app \\
121 & 151 & 0.003 & 0.007 & 0.007 & app \\
122 & 207 & 0.012 & 0.016 & 0.016 & app \\
126 & 312 & 0.026 * & 0.012 * & -0.014 ** & app \\
130 & 417 & -0.001 & -0.001 & -0.001 & app \\
172 & 489 & 0.023 ** & 0.025 ** & -0.003 & app \\
174 & 491 & 0.003 & -0.001 & -0.001 & app \\
255 & 525 & 0.001 & -0.003 & -0.003 & app \\
267 & 124 & 0.015 & 0.011 & 0.011 & app \\
\midrule
avg & 11 & 0.012 & 0.010 & 0.005 & \\
    \bottomrule
    \end{tabular}
    \caption{Average cost difference between pairs of K-approval votes and inferred K-approval votes. A positive score implies that the statistic for K-approval was higher.}
    \label{tab:pb cost shift approval inferred}
\end{table}

% \begin{table}[ht]
% \small
%     \centering
%     \begin{tabular}{lrrrr}
%     \toprule
%     {id} & n & top-1 & top-3 & avg \\
%     \midrule
% 103         &  193 &    0.05 &    0.07 &   0.07 \\
% 118         &   94 &    0.14 &    0.15 &   0.15 \\
% 119         &  127 &    0.06 &    0.13 &   0.13 \\
% 121         &  119 &    0.06 &    0.05 &   0.05 \\
% 122         &  158 &    0.08 &    0.11 &   0.11 \\
% 126         &  209 &    0.08 &    0.33 &  -0.42 \\
% 130         &  272 &    0.00 &    0.00 &  -0.01 \\
% 172         &  311 &    0.23 &    0.37 &  -0.09 \\
% 174         &  318 &    0.01 &   -0.00 &  -0.00 \\
% 255         &  332 &    0.02 &   -0.02 &  -0.02 \\
% 267         &  100 &    0.08 &    0.10 &   0.10 \\
% \midrule
% $> 0$& 11 & 10 & 8 & 6 \\
% $< 0$& 11 & 0 & 1 & 4 \\
%     \bottomrule
%     \end{tabular}
%     \caption{Net portion of voters who had a higher average cost statistic for their K-approval ballot than for their inferred K-approval ballot. A negative number implies that there was a net portion where the average cost was higher on their inferred K-approval ballot. The table describes the average project cost of the top-1, top-3 and all submitted projects on the ballot.}
%     \label{tab:pb cost net shift approval inferred}
% \end{table}

We can similarly compare K-approval votes with inferred K-approval votes to isolate any effect the question to rank the projects may have. To ensure individual-level comparability, we truncated the K-ranking ballots to match the number of projects selected by the voter at their equivalent K-approval ballot ($K_i$) to obtain the inferred K-approval ballot. We present the cost statistics in Tables~\ref{tab:pb cost shift approval inferred} and \ref{tab:pb cost net shift overview}.

We would have expected this effect to be small given the visual similarities in the interface: the K-ranking interface is essentially the same as K-approval, followed by an additional step where the voter is asked for their preferred order among the selected projects. Any effect would be attributed to the additional consideration step: this suggests that the step of asking voters to rank their project results in a different cost trade-off than immediately asking them for a smalller selection of projects. 

\subsection{Implicit constraints from voting methods}
\label{sec:pb implicit constraints}
\begin{table*}[ht]
% updated 240109: df_analysis_project_cost_ranking
\small
    \centering
    \begin{tabular}{p{0.04\textwidth}rrrrrrrrrrrrr}
    \toprule
    {} & app$_1$ & app$_2$ & app$_3$ & app$_4$ & app$_5$ & full$_v$ & full$_k$ & MK & K1$_v$ & K1$_k$ & kn$_p$ & kn$_s$ \\
    \midrule
top-1 &  1.22 &  1.20 &  1.19 &  1.19 &  1.17 &  1.16 &  1.16 & 1.16 & 1.13 & 1.19 & 1.14 &  1.0 \\
top-3 &  1.11 &  1.11 &  1.09 &  1.08 &  1.08 &  1.07 &  1.07 & 1.07 & 1.07 & 1.09 & 1.11 &  1.0 \\
avg  &  1.03 &  1.04 &  1.02 &  1.03 &  1.03 &  1.02 &  1.02 & 1.02 & 1.03 & 1.02 & 1.07 &  1.0 \\
    \bottomrule
    \end{tabular}    
    \caption{Average cost ratio of different aggregation methods applied to the same ballot. The ratio is the value divided by the same value for the knapsack (skip) aggregation when considering all selected projects, the 3 most expensive selected projects or the most expensive project. The column headers refer to the various aggregation methods.
    % app$_K$ is K-approval, full is a full Borda ranking, MK is Borda-MK, K1 is Borda-K1, $_v$ refers to voter-specific K's, and $_k$ to election-specific K's. kn$_p$ is knapsack with partial allocation and kn$_s$ with skip-allocation.
    }
    \label{tab:pb cost comparison aggregation ranking}
\end{table*}
\begin{table}[ht!]
\small
\small
    \centering
% updated 240109 df_costshift_appsim_knsim
\begin{tabular}{lrrrrl}
\toprule
    {id} & n & top-1 & top-3 & avg & Prim. \\
\midrule
16 & 681 & -0.011 ** & -0.012 ** & 0.0 & rank \\
93 & 1460 & -0.002 ** & -0.007 ** & 0.001 & rank \\
103 & 247 & 0.021 ** & 0.07 ** & 0.041 ** & app \\
106 & 100 & -0.008 ** & -0.011 ** & 0.005 ** & rank \\
118 & 119 & 0.077 ** & 0.064 ** & 0.034 ** & app \\
119 & 155 & 0.043 ** & 0.108 ** & 0.073 ** & app \\
121 & 151 & 0.022 ** & 0.062 ** & 0.032 ** & app \\
122 & 208 & 0.017 ** & 0.068 ** & 0.049 ** & app \\
126 & 314 & 0.028 ** & 0.016 ** & 0.008 ** & app \\
130 & 417 & 0.01 ** & 0.017 ** & -0.004 ** & app \\
152 & 228 & 0.073 ** & 0.065 ** & 0.02 ** & knap \\
171 & 103 & 0.0 & 0.0 & 0.0 & knap \\
172 & 491 & 0.0 & 0.001 ** & -0.0 & app \\
174 & 492 & 0.019 ** & 0.039 ** & 0.013 ** & app \\
186 & 240 & 0.0 & 0.0 & 0.0 & rank \\
194 & 139 & -0.028 * & 0.026 ** & 0.037 ** & knap \\
214 & 1747 & -0.008 ** & -0.005 ** & 0.004 ** & rank \\
225 & 440 & 0.0 & 0.0 & 0.0 & rank \\
242 & 378 & 0.107 ** & -0.011 & -0.011 & rank \\
243 & 304 & 0.056 ** & 0.098 ** & 0.042 ** & rank \\
244 & 1383 & 0.007 ** & -0.0 & 0.01 ** & rank \\
245 & 1285 & -0.012 ** & -0.01 ** & 0.0 & rank \\
247 & 854 & -0.005 ** & -0.01 ** & 0.0 & rank \\
248 & 109 & 0.007 & 0.041 ** & 0.029 ** & knap \\
249 & 356 & 0.003 & -0.007 ** & 0.007 ** & knap \\
250 & 411 & -0.017 ** & 0.064 ** & 0.059 ** & knap \\
251 & 128 & 0.044 ** & 0.046 ** & 0.031 ** & knap \\
255 & 525 & 0.135 ** & 0.055 ** & 0.016 ** & app \\
266 & 411 & -0.031 ** & -0.01 ** & 0.001 & rank \\
267 & 126 & 0.0 & 0.0 & 0.0 & app \\
270 & 425 & 0.0 & 0.0 & 0.0 & rank \\
\midrule
avg & 29 & 0.017 & 0.024 & 0.016 & \\
\bottomrule
\end{tabular} 
    \caption{Average cost difference between pairs of inferred 4-approval votes and inferred knapsack votes. A positive score implies that the statistic for inferred 4-approval was higher.}
    \label{tab:pb cost shift approval inferred knapsack inferred}
\end{table}

In order to isolate the effect of implicit constraints imposed by a voting method on the cost of selected projects, we compare the effect of a number of voting methods on the aggregated allocation. We report the average cost statistics of different ranking aggregation methods across elections in Table~\ref{tab:pb cost comparison aggregation ranking}, as ratio of the knapsack-skip aggregation. This shows very clearly that across elections, the standard knapsack aggregation results in lower-cost projects to be selected in aggregate, in line with the findings of Goel et al. 

To compare the effect of implicit constraints of the elicitation method (excluding the effect of aggregation), we infer both 4-approval and knapsack ballots from the same ranking votes, and report the cost statistics of this ballot pair in Tables~\ref{tab:pb cost shift approval inferred knapsack inferred} and \ref{tab:pb cost net shift overview}. We observe that the average cost is significantly higher in nearly all elections under 4-approval constraints than under knapsack constraints. This picture is more nuanced when we limit ourselves to the most expensive selected projects per voter, with significantly higher statistics for both ballots. This demonstrates that even if the projects are entirely independent, and if voters would exactly follow the same ranking to inform their approval and knapsack ballots, we would observe differences in the cost statistics.

\section{Abandonment rate and median time spent}
\begin{table*}[ht]
\small
    \centering
        \includegraphics[width=0.7\linewidth]{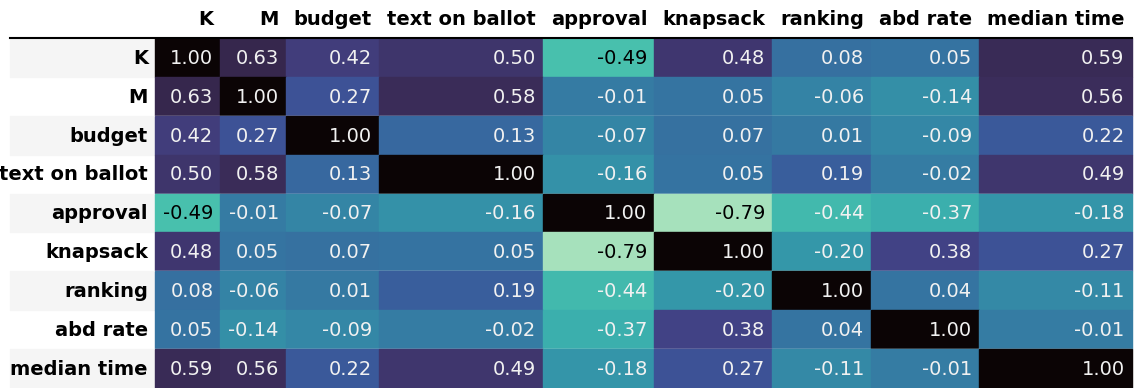}
%     \begin{tabular}{lrrp{0.05\textwidth}rrrrp{0.065\textwidth}p{0.065\textwidth}}
%     \toprule
%     {} & K & M & ballot length & budget & approval & knapsack & ranking & abd rate & median time \\
%     \midrule
% K & 1.00 & 0.63 & 0.49 & 0.43 & -0.48 & 0.49 & 0.05 & 0.05 & 0.58 \\
% M & 0.63 & 1.00 & 0.58 & 0.27 & -0.01 & 0.05 & -0.06 & -0.15 & 0.56 \\
% ballot length & 0.49 & 0.58 & 1.00 & 0.13 & -0.16 & 0.05 & 0.19 & -0.00 & 0.48 \\
% budget & 0.43 & 0.27 & 0.13 & 1.00 & -0.07 & 0.07 & 0.01 & -0.09 & 0.22 \\
% approval & -0.48 & -0.01 & -0.16 & -0.07 & 1.00 & -0.79 & -0.44 & -0.35 & -0.18 \\
% knapsack & 0.49 & 0.05 & 0.05 & 0.07 & -0.79 & 1.00 & -0.20 & 0.36 & 0.27 \\
% ranking & 0.05 & -0.06 & 0.19 & 0.01 & -0.44 & -0.20 & 1.00 & 0.04 & -0.11 \\
% abd rate & 0.05 & -0.15 & -0.00 & -0.09 & -0.35 & 0.36 & 0.04 & 1.00 & -0.02 \\
% median time & 0.58 & 0.56 & 0.48 & 0.22 & -0.18 & 0.27 & -0.11 & -0.02 & 1.00 \\
%     \bottomrule
%     \end{tabular}
        \caption{Correlations of election characteristics and abandonment rate, median time. K is defined as the largest number of projects selected by any participant, and M is the number of projects available to choose from. (n = 121) Colors indicate magnitude.}
    \label{tab:pb election correlations}
\end{table*}

\begin{table*}[t]
\small
    \centering
    \begin{tabular}{l|cc|cc}
        \toprule
        Variable & \multicolumn{2}{c}{Median Time Spent} & \multicolumn{2}{c}{Abandonment Rate} \\
        {} & coefficient & 95\% confidence interval & coefficient & 95\% confidence interval \\
        \midrule
        K & 4.7 * &  0.3 -- 9.0                     & -0.0012 &  -0.005 -- 0.002 \\
        M & 2.6 * & 0.3 -- 4.9                      & -0.0010 &  -0.003 -- 0.001 \\
        text on the ballot & 0.005 * & 0.001 -- 0.009    & 0.000 & -0.000 -- 0.000 \\
        budget & 0.000 & 0.000 -- 0.000             & 0.000 & -0.000 -- 0.000 \\
        approval method & 20 * & 5 -- 36            & -0.0012 & -0.01 -- 0.01 \\
        knapsack method & 33 ** & 13 -- 53          & 0.043 *** & 0.027 -- 0.059 \\ 
        ranking method & -10 & -34 -- 15            & 0.014 & -0.005 -- 0.034 \\
         \bottomrule
    \end{tabular}
    \caption{Multiple Linear Regression (121 elections). K is the largest number of projects selected in any submitted vote. M is the number of available projects on the ballot. Text on the ballot is the cumulative number of characters in project descriptions in English. Approval, knapsack, and ranking are binary indicators of using the respective method.}
    \label{tab:pb multiple linear regression}
\end{table*}

\begin{figure}[t]
\centering
    \includegraphics[width=0.7\linewidth]{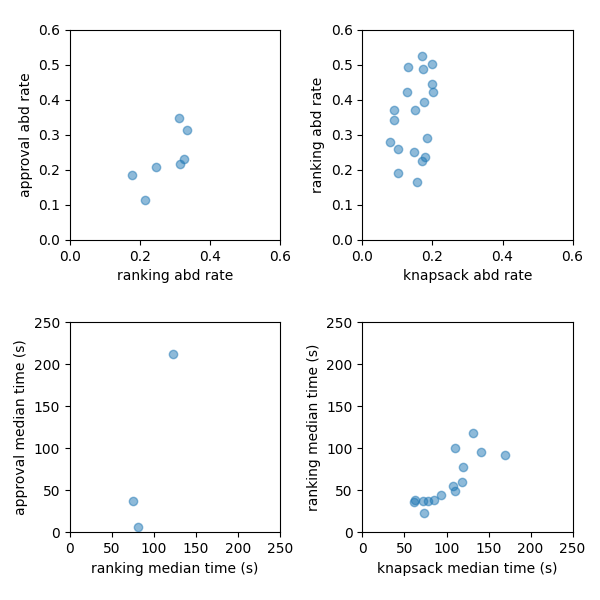}
    \caption{Abandonment (abd) rate and median time for secondary K-approval, K-ranking and knapsack ballots at the same election. Voters were randomly assigned to either secondary voting method.}
    \label{fig:pb abandonment median time}
\end{figure}
% \begin{figure}
%     \includegraphics[width=0.95\linewidth]{figs/abandonment_approval_ranking.png}
%     \caption{Abandonment rate for secondary K-approval and K-ranking ballots at the same election. Voters were randomly assigned to either secondary voting method.}
%     \label{fig:pb abandonment approval ranking}
% \end{figure}
% \begin{figure}
%     \includegraphics[width=0.95\linewidth]{figs/abandonment_ranking_knapsack.png}
%     \caption{Abandonment rate for secondary K-ranking and knapsack ballots at the same election. Voters were randomly assigned to either secondary voting method.}
%     \label{fig:pb abandonment ranking knapsack}
% \end{figure}
% \begin{figure}
%     \includegraphics[width=0.95\linewidth]{figs/median_time_ranking_knapsack.png}
%     \caption{Median time spent for secondary K-ranking and knapsack ballots at the same election. Voters were randomly assigned to either secondary voting method.}
%     \label{fig:pb median time ranking knapsack}
% \end{figure}

Election organizers indicated in private communications that they choose the K-approval voting method for its perceived ease of use for voters. We gauge voter usability by analyzing the abandonment rate and median time spent during ballot completion. We present in Table~\ref{tab:pb election correlations} the correlation of the main ballot design choices (number of projects to select $K$, number of projects on the ballot $M$, text on the ballot, budget available, voting method) with each other and the abandonment rate and median time spent for 121 ballots. In Table~\ref{tab:pb multiple linear regression} we present the results of multiple-linear regressions for both the median time spent and the abandonment rate\footnote{Abandonment rate is only available in aggregate. Time spent is anonymized at the per-voter level to a percentile score.} on the ballot design choices.

As anticipated, choices that affect ballot complexity are all significantly predictive for a higher median time spent, but do not predict a higher abandonment rate. The knapsack voting method is the only choice that significantly predicts a higher abandonment rate. However, as it is not possible to have no voting method at all, we should note that the confidence intervals for the different voting methods do overlap. Therefore, we can confidently conclude that voters in elections with knapsack voting spend more time and are a little more likely to exit the ballot than voters in elections that use the approval method. However, we can not assert the same when we compare voters in elections using the knapsack and ranking methods -- the more relevant comparison, as the knapsack ballot can be inferred from ranking.

These findings only hint at associations, and causality cannot be inferred. For instance, cities choosing knapsack voting might employ a more innovative or extensive recruitment than those choosing approval voting, leading to higher dropout rates. To establish causality, randomized controlled trials would be needed. For some elections, we have effectively established exactly that, by randomly assigning a secondary voting method to voters. Unfortunately, the number of elections where we have sufficient votes available in two randomly assigned voting methods, is limited. In Figure~\ref{fig:pb abandonment median time} we present the randomized controlled data that we have available in our data set for the approval-ranking and the ranking-knapsack ballot combinations. These data points represent secondary ballots where voters were (after completing the same primary ballot) randomly assigned to different secondary ballots (e.g. after a primary approval ballot, they were assigned to either a knapsack or a 5-ranking secondary ballot). This data shows that the abandonment rate for knapsack ballots is actually lower than K-ranking, even though the median time spent is higher. 

This effect could be partly attributed to the fact that the K-ranking and K-approval interfaces are so similar, and that a K-ranking secondary ballot always followed a K-approval primary ballot, and vice versa. This could reinforce a higher median time for knapsack (more familiarity between ranking and approval) and higher abandonment rate for ranking and approval due to voter confusion. Consequently, these findings cannot be readily generalized to primary voting methods without further experimentation. 

In summary, the question whether knapsack voting results in a higher abandonment rate remains inconclusive, and further research is warranted. 

% \section{Required sample size}
% \lodewijk{This subsection needs further analysis and needs to be fleshed out. It has some parallels in goal with Garg et al (2019).}

% \begin{enumerate}
%     \item How many people do we need to participate in an election to be confident that we have a large enough sample size? With our data, we can infer this:
%     \item For a given election, we draw X samples with replacement (bootstrap) from the actual voters in that election
%     \item We aggregate the results for each of these bootstrapped elections
%     \item We calculate the expected distance between a bootstrapped election aggregate and the actual aggregate
%     \item What is the smallest response number of voters m that we need to be 95\% confident that we get the complete picture?
%     \item We calculate this for each election. Then we plot this against several factors.
%     \item We then infer ballots from rankings and insert those into the plot as well. 
% \end{enumerate}

\section{Discussions and conclusions}
\label{sec:pb discussion}

With this analysis, we have uncovered valuable insights that can aid PB election organizers. By examining ballot pairs from the same voters and leveraging long K-ranking votes, we made the following observations regarding voting methods and their effect on the average cost of selected projects:
\begin{enumerate}
% Validity
    \item Secondary ballots serve as effective tools for comparing voting methods.
    \item For available K-ranking data, the specific Borda count variant chosen has a limited effect on project allocation. The choice of $K$ in K-approval voting has a bigger impact.
    \item Inferring knapsack votes from sufficiently long K-ranking votes is a valid approximation method when no digital interface is available.
\end{enumerate}
We also provide a more rigorous and nuanced understanding of the following insights in existing literature about the average cost of selected projects under different voting methods:
\begin{enumerate}
\setcounter{enumi}{3}
% Project cost
    \item In most elections, voters tend to select significantly more expensive projects using the K-approval method, compared to the knapsack method. 
    However, when we limit the analysis to the most expensive projects selected by a voter, they select significantly higher-cost projects under either method, depending on the election. 
    % (See Table~\ref{tab:pb cost shift approval knapsack})
    \item In several elections, voters tend to select significantly less expensive projects using the knapsack interface, than when the same constraints are applied to their vote cast with the K-ranking interface. Explicit budget constraints contribute to the lower-cost effect of the knapsack voting method. 
    % (See Table~\ref{tab:pb cost shift knapsack inferred})
    \item In some K-ranking ballots, voters ranked their most expensive projects lower than would be expected based on their K-approval votes. 
    % (See Table~\ref{tab:pb cost shift approval inferred})
    \item In most elections, voters would select less expensive projects if their choice was budget constrained, than by number of projects. This excludes effects from the voting interface or information presented. The implicit constraints of the voting method itself constribute to the lower-cost effect of the knapsack voting method. 
    % (See Table~\ref{tab:pb cost shift approval inferred knapsack inferred})
    \item Aggregating ranking ballots with a knapsack method results in selecting less expensive projects compared to aggregation with K-approval or Borda Count. 
\end{enumerate}
These findings corroborate and expand upon results from \citep{Goel2019}: knapsack voting tends to result in lower-cost selected budgets. We find that it also results in lower-cost individual ballots. This cost effect is influenced by both explicit constraints and presented information, as well as the implicit constraint of the method itself. Notably, this cost effect is consistent across all selected projects, but not for the most expensive selected projects. This suggests that voters prioritize cost considerations for cheaper projects, which should be confirmed in more qualitative user studies. 

The conclusion that explicit constraints matter, implies that while knapsack ballots and inferred knapsack ballots are meaningfully similar, and ranked ballot elicitation is a viable alternative to knapsack elicitation (for example, on paper), it is not a perfect substitute and may impact the selected projects. However, since the cost effects align, this does not necessarily have to be a problem. 
We examined abandonment rates and median time spent for primary ballots, revealing:
\begin{enumerate}
\setcounter{enumi}{8}
% Difficulty of the ballot
    \item Ballot design can affect the effort that voters have to spend on a ballot. Ballot complexity (e.g. number of projects to choose from, number of projects to select, and text on the ballot) positively correlated with the median time spent by voters. However, there is no significant correlation with the abandonment rate. 
    \item Elections using the knapsack voting method correlate with a higher median time spent and higher abandonment rate. Still, the cause remains unclear and the causality remains inconclusive. 
\end{enumerate}
A lot more analysis is possible with this data set, and we hope that this will provide a fruitful avenue for other researchers to test various hypotheses. 

This study demonstrates the feasibility and value of comparing voting methods in practical settings, although this comes with its own limitations. While it allows a direct evaluation of real voter considerations under different elicitation methods, it does not allow a fully randomized treatment where within the same election. We would encourage more experimental ballots to be included in real PB processes, in order for these findings to be evaluated and understood more thoroughly, and sharing anonymized data of such ballots. Only that way, the potential confounding effect of various factors such as turnout and recruitment method can be adaquately addressed. 

Future work could also include surveys and user studies with real voters in various processes that help understand to what extent voters are aware of various specifics of the voting process (e.g. the aggregation method selected), and whether they considered this in their voting strategy. This could more definitively inform assumptions in the aggregation method. 

To really understand whether knapsack voting is a harder voting method for voters and whether that deters people, we will need to design dedicated experiments with real voters on real issues. Analysis of secondary ballot data (such as the data we are releasing) is a promising avenue to develop such insights, but will require data from more elections for answering some of the questions.

\subsection{Impact considerations}
The cleaning and anonymization process of the dataset has been executed to maximally preserve reproducibility while assuring privacy of the participants. Cities have not been anonymized as these processes were public. Conclusions were framed to improve understanding for practitioners.

\anonymize{\section{Acknowledgments}
The authors acknowledge the many PB organizers that agreed to include a research ballot as part of their election. Part of the data collection efforts was funded through grants from Cisco Systems [unrestricted donation] and from the US Department of the Navy for 'Collaborative Decision Making at Scale: Bridging Theory and Practice' [grant number N00014-15-1-2786]. We thank Sukolsak Sakshuwong for developing and maintaining the Stanford Participatory Budgeting platform for many years, and thank Tanja Aitamurto and Anilesh K. Krishnaswamy for their contributions. We appreciate the helpful discussions with and feedback during the data collection design from Nikhil Garg, the analysis support from Megan Mou and further feedback on the paper from Mohak Goyal, Irene Lo and anonymous reviewers. 

% Chat GPT was used to rewrite sections of this paper and condense it. All editorial decisions were manually implemented by authors. 
}{}
\bibliographystyle{aaai22}
\bibliography{refs}

\begin{thebibliography}{38}
\providecommand{\natexlab}[1]{#1}

\bibitem[{Arana-Catania et~al.(2021)Arana-Catania, Lier, Procter, Tkachenko,
  He, Zubiaga, and Liakata}]{arana-catania_citizen_2021}
Arana-Catania, M.; Lier, F.-A.~V.; Procter, R.; Tkachenko, N.; He, Y.; Zubiaga,
  A.; and Liakata, M. 2021.
\newblock Citizen {Participation} and {Machine} {Learning} for a {Better}
  {Democracy}.
\newblock \emph{Digital Government: Research and Practice}, 2(3): 1--22.

\bibitem[{Arnstein(1969)}]{arnstein_ladder_1969}
Arnstein, S.~R. 1969.
\newblock A {Ladder} {Of} {Citizen} {Participation}.
\newblock \emph{Journal of the American Institute of Planners}, 35(4):
  216--224.

\bibitem[{Aziz and Shah(2021)}]{rudas_participatory_2021}
Aziz, H.; and Shah, N. 2021.
\newblock Participatory {Budgeting}: {Models} and {Approaches}.
\newblock In Rudas, T.; and Péli, G., eds., \emph{Pathways {Between} {Social}
  {Science} and {Computational} {Social} {Science}}, 215--236. Cham: Springer
  International Publishing.
\newblock ISBN 978-3-030-54935-0 978-3-030-54936-7.
\newblock Series Title: Computational Social Sciences.

\bibitem[{Bartocci et~al.(2022)Bartocci, Grossi, Mauro, and
  Ebdon}]{bartocci_journey_2022}
Bartocci, L.; Grossi, G.; Mauro, S.~G.; and Ebdon, C. 2022.
\newblock The journey of participatory budgeting: a systematic literature
  review and future research directions.
\newblock \emph{International Review of Administrative Sciences},
  002085232210789.

\bibitem[{Benadè et~al.(2017)Benadè, Nath, Procaccia, and
  Shah}]{benade_preference_2017}
Benadè, G.; Nath, S.; Procaccia, A.~D.; and Shah, N. 2017.
\newblock Preference {Elicitation} {For} {Participatory} {Budgeting}.
\newblock \emph{Journal of the ACM}, 1(1): 27.

\bibitem[{Callahan(2007{\natexlab{a}})}]{callahan_citizen_2007}
Callahan, K. 2007{\natexlab{a}}.
\newblock Citizen {Participation}: {Models} and {Methods}.
\newblock \emph{International Journal of Public Administration}, 30(11):
  1179--1196.

\bibitem[{Callahan(2007{\natexlab{b}})}]{callahan_elements_2007}
Callahan, K., ed. 2007{\natexlab{b}}.
\newblock \emph{Elements of effective governance: measurement, accountability
  and participation}.
\newblock Number 126 in Public administration and public policy. Boca Raton:
  CRC/Taylor \& Francis.
\newblock ISBN 978-0-8493-7096-0.
\newblock OCLC: ocm69013806.

\bibitem[{Cantador and Cortés-Cediel(2018)}]{cantador_towards_2018}
Cantador, I.; and Cortés-Cediel, M.~E. 2018.
\newblock Towards increasing citizen engagement in participatory budgeting
  digital tools.
\newblock In \emph{Proceedings of the 19th {Annual} {International}
  {Conference} on {Digital} {Government} {Research}}, 1--2. ACM Press.
\newblock ISBN 978-1-4503-6526-0.

\bibitem[{de~Borda(1784)}]{de_borda_memoire_1784}
de~Borda, J.-C. 1784.
\newblock Mémoire sur les élections au scrutin.
\newblock \emph{Histoire de l'Académie royale des sciences}, Année 1781:
  31--34.

\bibitem[{Ebdon and Franklin(2006)}]{ebdon_citizen_2006}
Ebdon, C.; and Franklin, A.~L. 2006.
\newblock Citizen {Participation} in {Budgeting} {Theory}.
\newblock \emph{Public Administration Review}, 66(3): 437--447.

\bibitem[{Fain, Goel, and Munagala(2016)}]{fain_core_2016}
Fain, B.; Goel, A.; and Munagala, K. 2016.
\newblock The {Core} of the {Participatory} {Budgeting} {Problem}.
\newblock In Cai, Y.; and Vetta, A., eds., \emph{Web and {Internet}
  {Economics}}, volume 10123, 384--399. Berlin, Heidelberg: Springer Berlin
  Heidelberg.
\newblock ISBN 978-3-662-54109-8 978-3-662-54110-4.
\newblock Series Title: Lecture Notes in Computer Science.

\bibitem[{Fairstein, Benade, and Gal(2023)}]{fairstein_participatory_2023}
Fairstein, R.; Benade, G.; and Gal, K. 2023.
\newblock Participatory {Budgeting} {Design} for the {Real} {World}.
\newblock In \emph{Proceedings of the {AAAI} {Conference} on {Artificial}
  {Intelligence}}, volume~37.

\bibitem[{Falanga and Lüchmann(2020)}]{falanga_participatory_2020}
Falanga, R.; and Lüchmann, L. H.~H. 2020.
\newblock Participatory budgets in {Brazil} and {Portugal}: comparing patterns
  of dissemination.
\newblock \emph{Policy Studies}, 41(6): 603--622.

\bibitem[{Fung(2006)}]{fung_varieties_2006}
Fung, A. 2006.
\newblock Varieties of {Participation} in {Complex} {Governance}.
\newblock \emph{Public Administration Review}, 66(s1): 66--75.

\bibitem[{Garg et~al.(2019)Garg, Gelauff, Sakshuwong, and Goel}]{garg_who_2019}
Garg, N.; Gelauff, L.; Sakshuwong, S.; and Goel, A. 2019.
\newblock Who is in {Your} {Top} {Three}?
\newblock In \emph{{HCOMP} 2019}.

\bibitem[{Gelauff and Goel(2024)}]{gelauff_participatory_2024}
Gelauff, L.; and Goel, A. 2024.
\newblock Participatory {Budgeting} {Preferences} {Data} {Set}.

\bibitem[{Gelauff et~al.(2020)Gelauff, Goel, Munagala, and
  Yandamuri}]{gelauff_advertising_2020}
Gelauff, L.; Goel, A.; Munagala, K.; and Yandamuri, S. 2020.
\newblock Advertising for {Demographically} {Fair} {Outcomes}.
\newblock In \emph{{ArXiv}, Presented at: 4th Workshop on Mechanism Design for
  Social Good}. Online.
\newblock ArXiv: 2006.03983.

\bibitem[{Goel et~al.(2019)Goel, Krishnaswamy, Sakshuwong, and
  Aitamurto}]{Goel2019}
Goel, A.; Krishnaswamy, A.~K.; Sakshuwong, S.; and Aitamurto, T. 2019.
\newblock Knapsack {Voting} for {Participatory} {Budgeting}.
\newblock \emph{ACM Transactions on Economics and Computation}, 7(2): 1--27.

\bibitem[{Goyal et~al.(2023)Goyal, Sakshuwong, Sarmasarkar, and
  Goel}]{goyal_lowsample_2023}
Goyal, M.; Sakshuwong, S.; Sarmasarkar, S.; and Goel, A. 2023.
\newblock {Low Sample Complexity Participatory Budgeting}.
\newblock In Etessami, K.; Feige, U.; and Puppis, G., eds., \emph{50th
  International Colloquium on Automata, Languages, and Programming (ICALP
  2023)}, volume 261 of \emph{Leibniz International Proceedings in Informatics
  (LIPIcs)}, 70:1--70:20. Dagstuhl, Germany: Schloss Dagstuhl --
  Leibniz-Zentrum f{\"u}r Informatik.
\newblock ISBN 978-3-95977-278-5.

\bibitem[{Hagelskamp et~al.(2016)Hagelskamp, Rinehart, Silliman, and
  Schleifer}]{hagelskamp_public_2016}
Hagelskamp, C.; Rinehart, C.; Silliman, R.; and Schleifer, D. 2016.
\newblock Public {Spending}, by the {People}. {Participatory} {Budgeting} in
  the {United} {States} and {Canada} in 2014-2015.
\newblock Technical report, Public Agenda.

\bibitem[{Holston, Issarny, and Parra(2016)}]{holston_engineering_2016}
Holston, J.; Issarny, V.; and Parra, C. 2016.
\newblock Engineering software assemblies for participatory democracy: the
  participatory budgeting use case.
\newblock In \emph{Proceedings of the 38th {International} {Conference} on
  {Software} {Engineering} {Companion}}, 573--582. Austin Texas: ACM.
\newblock ISBN 978-1-4503-4205-6.

\bibitem[{Johnson, Carlson, and Reynolds(2021)}]{johnson_testing_2021}
Johnson, C.; Carlson, H.~J.; and Reynolds, S. 2021.
\newblock Testing the {Participation} {Hypothesis}: {Evidence} from
  {Participatory} {Budgeting}.
\newblock \emph{Political Behavior}, 45: 3--32.

\bibitem[{Kim et~al.(2016)Kim, Jung, Ko, Han, Lee, Kim, and
  Kim}]{kim_budgetmap_2016}
Kim, N.~W.; Jung, J.; Ko, E.-Y.; Han, S.; Lee, C.~W.; Kim, J.; and Kim, J.
  2016.
\newblock {BudgetMap}: {Engaging} {Taxpayers} in the {Issue}-{Driven}
  {Classification} of a {Government} {Budget}.
\newblock In \emph{Proceedings of the 19th {ACM} {Conference} on
  {Computer}-{Supported} {Cooperative} {Work} \& {Social} {Computing}},
  1028--1039. San Francisco California USA: ACM.
\newblock ISBN 978-1-4503-3592-8.

\bibitem[{Langton(1979)}]{langton_american_1979}
Langton, S. 1979.
\newblock American citizen participation: {A} deep-rooted tradition.
\newblock \emph{National Civic Review}, 68(8): 403--422.

\bibitem[{Lerner and Pape(2020)}]{lerner_budgeting_2020}
Lerner, J.; and Pape, M. 2020.
\newblock Budgeting for {Equity}: {How} {Can} {Participatory} {Budgeting}
  {Advance} {Equity} in the {United} {States}?
\newblock \emph{Journal of Deliberative Democracy}, 12(2).

\bibitem[{Miller, Hildreth, and Stewart(2019)}]{miller_modes_2019}
Miller, S.~A.; Hildreth, R.~W.; and Stewart, L.~M. 2019.
\newblock The {Modes} of {Participation}: {A} {Revised} {Frame} for
  {Identifying} and {Analyzing} {Participatory} {Budgeting} {Practices}.
\newblock \emph{Administration \& Society}, 51(8): 1254--1281.

\bibitem[{Munagala, Shen, and Wang(2022)}]{munagala_auditing_2022}
Munagala, K.; Shen, Y.; and Wang, K. 2022.
\newblock Auditing for {Core} {Stability} in {Participatory} {Budgeting}.
\newblock In Hansen, K.~A.; Liu, T.~X.; and Malekian, A., eds., \emph{Web and
  {Internet} {Economics}}, volume 13778, 292--310. Cham: Springer International
  Publishing.
\newblock ISBN 978-3-031-22831-5 978-3-031-22832-2.
\newblock Series Title: Lecture Notes in Computer Science.

\bibitem[{Peters, Pierczyński, and Skowron(2021)}]{peters_proportional_2021}
Peters, D.; Pierczyński, G.; and Skowron, P. 2021.
\newblock Proportional {Participatory} {Budgeting} with {Additive} {Utilities}.
\newblock In \emph{Advances in {Neural} {Information} {Processing} {Systems} 34
  ({NeurIPS} 2021)}, volume~34, 12726--12737. Curran Associates, Inc.

\bibitem[{Pina et~al.(2022)Pina, Torres, Royo, and
  Garcia-Rayado}]{pina_decide_2022}
Pina, V.; Torres, L.; Royo, S.; and Garcia-Rayado, J. 2022.
\newblock Decide {Madrid}: {A} {Spanish} best practice on e-participation.
\newblock In Randma-Liiv, T.; and Lember, V., eds., \emph{Engaging {Citizens}
  in {Policy} {Making}}. Edward Elgar Publishing.
\newblock ISBN 978-1-80037-436-2 978-1-80037-435-5.

\bibitem[{Rey and Maly(2023)}]{rey_computational_2023}
Rey, S.; and Maly, J. 2023.
\newblock The ({Computational}) {Social} {Choice} {Take} on {Indivisible}
  {Participatory} {Budgeting}.
\newblock ArXiv:2303.00621.

\bibitem[{Serramia et~al.(2019)Serramia, Lopez-Sanchez, Rodriguez-Aguilar, and
  Escobar}]{serramia_optimising_2019}
Serramia, M.; Lopez-Sanchez, M.; Rodriguez-Aguilar, J.~A.; and Escobar, P.
  2019.
\newblock Optimising participatory budget allocation: the {Decidim} use case.
\newblock \emph{Artificial Intelligence Research and Development}, 319:
  193--202.

\bibitem[{Sintomer, Herzberg, and Röcke(2008)}]{sintomer_participatory_2008}
Sintomer, Y.; Herzberg, C.; and Röcke, A. 2008.
\newblock Participatory {Budgeting} in {Europe}: {Potentials} and {Challenges}:
  {Participatory} budgeting in {Europe}.
\newblock \emph{International Journal of Urban and Regional Research}, 32(1):
  164--178.

\bibitem[{Society(2016)}]{the_democratic_society_digital_2016}
Society, T.~D. 2016.
\newblock Digital tools and {Scotland}'s {Participatory} {Budgeting} programme.
\newblock Technical report, The Democratic Society.

\bibitem[{Stolicki, Szufa, and Talmon(2020)}]{stolicki_pabulib_2020}
Stolicki, D.; Szufa, S.; and Talmon, N. 2020.
\newblock Pabulib: {A} {Participatory} {Budgeting} {Library}.
\newblock ArXiv:2012.06539.

\bibitem[{Su(2017)}]{su_porto_2017}
Su, C. 2017.
\newblock From {Porto} {Alegre} to {New} {York} {City}: {Participatory}
  {Budgeting} and {Democracy}.
\newblock \emph{New Political Science}, 39(1): 67--75.

\bibitem[{van Dijk(2012)}]{van_dijk_digital_2012}
van Dijk, J.~A. 2012.
\newblock Digital {Democracy}: {Vision} and {Reality}.
\newblock In Snellen, I. T.~M.; Thaens, M.; and Donk, W. B. H. J. v.~d., eds.,
  \emph{Public administration in the information age: revisited}, number volume
  19 in Innovation and the public sector, 49--62. Amsterdam ; Washington, DC:
  IOS Press.
\newblock ISBN 978-1-61499-136-6.

\bibitem[{Wang(2001)}]{wang_assessing_2001}
Wang, X. 2001.
\newblock Assessing {Public} {Participation} in {U}.{S}. {Cities}.
\newblock \emph{Public Performance \& Management Review}, 24(4): 322.

\bibitem[{Williams, Denny, and Bristow(2017)}]{williams_participatory_2017}
Williams, E.; Denny, E.~S.; and Bristow, D. 2017.
\newblock Participatory {Budgeting}: {An} {Evidence} {Review}.
\newblock Technical report, Public Policy Institute for Wales.

\end{thebibliography}

\clearpage
\appendix
\section{Appendix}
\begin{table*}[htb]
    % update 240109 cosine
    \small
    \centering
    \begin{tabular}{p{0.04\textwidth}rrrrrrrrrrrr}
    \toprule
    {} &  app$_1$ &  app$_2$ &  app$_3$ &  app$_4$ &  app$_5$ &  full$_v$ &  full$_k$ &  MK &  K1$_v$ &  K1$_k$ &  kn$_p$ &  kn$_s$ \\
    \midrule
    app$_1$ &  1.00 &  0.91 &  0.84 &  0.79 &  0.76 &  0.81 &  0.83 & 0.83 & 0.78 & 0.84 & 0.83 & 0.71 \\
    app$_2$ &  0.91 &  1.00 &  0.93 &  0.88 &  0.84 &  0.90 &  0.91 & 0.92 & 0.87 & 0.93 & 0.89 & 0.79 \\
    app$_3$ &  0.84 &  0.93 &  1.00 &  0.93 &  0.88 &  0.97 &  0.98 & 0.98 & 0.91 & 1.00 & 0.90 & 0.85 \\
    app$_4$ &  0.79 &  0.88 &  0.93 &  1.00 &  0.93 &  0.96 &  0.94 & 0.94 & 0.93 & 0.93 & 0.87 & 0.84 \\
    app$_5$ &  0.76 &  0.84 &  0.88 &  0.93 &  1.00 &  0.91 &  0.90 & 0.90 & 0.89 & 0.88 & 0.87 & 0.81 \\
    full$_v$ &  0.81 &  0.90 &  0.97 &  0.96 &  0.91 &  1.00 &  0.98 & 0.98 & 0.91 & 0.97 & 0.90 & 0.88 \\
    full$_k$ &  0.83 &  0.91 &  0.98 &  0.94 &  0.90 &  0.98 &  1.00 & 1.00 & 0.90 & 0.98 & 0.89 & 0.86 \\
    MK   &  0.83 &  0.92 &  0.98 &  0.94 &  0.90 &  0.98 &  1.00 & 1.00 & 0.90 & 0.98 & 0.90 & 0.87 \\
    K1$_v$  &  0.78 &  0.87 &  0.91 &  0.93 &  0.89 &  0.91 &  0.90 & 0.90 & 1.00 & 0.92 & 0.85 & 0.85 \\
    K1$_k$  &  0.84 &  0.93 &  1.00 &  0.93 &  0.88 &  0.97 &  0.98 & 0.98 & 0.92 & 1.00 & 0.91 & 0.85 \\
    kn$_p$  &  0.83 &  0.89 &  0.90 &  0.87 &  0.87 &  0.90 &  0.89 & 0.90 & 0.85 & 0.91 & 1.00 & 0.81 \\
    kn$_s$  &  0.71 &  0.79 &  0.85 &  0.84 &  0.81 &  0.88 &  0.86 & 0.87 & 0.85 & 0.85 & 0.81 & 1.00 \\
    \bottomrule
    \end{tabular}
    \caption{Cosine similarity between the proportion of the budget allocated to all projects under different aggregation methods, inferred from identical K-ranking ballots: K-approval (app$_K$), full Borda ranking, Borda-MK, Borda-K1 (with voter- ($_v$) and election-specific $K$ ($_k$)), knapsack with partial (kn$_p$) and skip-allocation (kn$_s$). Unformatted version of Table~\ref{tab:pb cosine ranking aggregations}}
\label{tab:pb cosine ranking aggregations unformatted}
\end{table*}

\begin{table*}[ht]
\small
    \centering
    \begin{tabular}{lrrp{0.05\textwidth}rrrrp{0.065\textwidth}p{0.065\textwidth}}
    \toprule
    {} & K & M & ballot text & budget & approval & knapsack & ranking & abd rate & median time \\
    \midrule
K & 1.00 & 0.63 & 0.49 & 0.43 & -0.48 & 0.49 & 0.05 & 0.05 & 0.58 \\
M & 0.63 & 1.00 & 0.58 & 0.27 & -0.01 & 0.05 & -0.06 & -0.15 & 0.56 \\
ballot text & 0.49 & 0.58 & 1.00 & 0.13 & -0.16 & 0.05 & 0.19 & -0.00 & 0.48 \\
budget & 0.43 & 0.27 & 0.13 & 1.00 & -0.07 & 0.07 & 0.01 & -0.09 & 0.22 \\
approval & -0.48 & -0.01 & -0.16 & -0.07 & 1.00 & -0.79 & -0.44 & -0.35 & -0.18 \\
knapsack & 0.49 & 0.05 & 0.05 & 0.07 & -0.79 & 1.00 & -0.20 & 0.36 & 0.27 \\
ranking & 0.05 & -0.06 & 0.19 & 0.01 & -0.44 & -0.20 & 1.00 & 0.04 & -0.11 \\
abd rate & 0.05 & -0.15 & -0.00 & -0.09 & -0.35 & 0.36 & 0.04 & 1.00 & -0.02 \\
median time & 0.58 & 0.56 & 0.48 & 0.22 & -0.18 & 0.27 & -0.11 & -0.02 & 1.00 \\
    \bottomrule
    \end{tabular}
        \caption{Correlations of election characteristics and abandonment rate, median time. K is defined as the largest number of projects selected by any participant, and M is the number of projects available to choose from. (n = 121) Unformatted version of Table~\ref{tab:pb election correlations}}
    \label{tab:pb election correlations unformatted}
\end{table*}

\end{document}